\pdfoutput=1
\PassOptionsToPackage{unicode=true}{hyperref} 
\PassOptionsToPackage{hyphens}{url}
\documentclass[]{sig-alternate-05-2015}
\usepackage{lmodern}
\usepackage{amssymb,amsmath}
\usepackage{ifxetex,ifluatex}
\usepackage{fixltx2e} 
\ifnum 0\ifxetex 1\fi\ifluatex 1\fi=0 
  \usepackage[T1]{fontenc}
  \usepackage[utf8]{inputenc}
  \usepackage{textcomp} 
\else 
  \usepackage{unicode-math}
  \defaultfontfeatures{Ligatures=TeX,Scale=MatchLowercase}
\fi
\IfFileExists{upquote.sty}{\usepackage{upquote}}{}
\IfFileExists{microtype.sty}{%
\usepackage[]{microtype}
\UseMicrotypeSet[protrusion]{basicmath} 
}{}
\IfFileExists{parskip.sty}{%
\usepackage{parskip}
}{
\setlength{\parindent}{0pt}
\setlength{\parskip}{6pt plus 2pt minus 1pt}
}
\usepackage{hyperref}
\hypersetup{
            pdftitle={LazyFP: Leaking FPU Register State using Microarchitectural Side-Channels},
            pdfborder={0 0 0},
            breaklinks=true}
\ifx\paragraph\undefined\else
\let\oldparagraph\paragraph
\renewcommand{\paragraph}[1]{\oldparagraph{#1}\mbox{}}
\fi
\ifx\subparagraph\undefined\else
\let\oldsubparagraph\subparagraph
\renewcommand{\subparagraph}[1]{\oldsubparagraph{#1}\mbox{}}
\fi

\makeatletter
\def\fps@figure{htbp}
\makeatother

\usepackage{booktabs}
\usepackage{listings}
\usepackage{xcolor}
\setcopyright{rightsretained}
\numberofauthors{2}
\author{
\alignauthor
Julian Stecklina\\
       \affaddr{Amazon Development Center Germany GmbH}\\
       \email{jsteckli@amazon.de}
\alignauthor
Thomas Prescher\\
       \affaddr{Cyberus Technology GmbH}\\
       \email{thomas.prescher@cyberus-technology.de}
}

\title{LazyFP: Leaking FPU Register State using Microarchitectural
Side-Channels}
\date{}

\begin{document}
\maketitle
\begin{abstract}
Modern processors utilize an increasingly large register set to
facilitate efficient floating point and SIMD computation. This large
register set is a burden for operating systems, as its content needs to
be saved and restored when the operating system context switches between
tasks. As an optimization, the operating system can defer the context
switch of the FPU and SIMD register set until the first instruction is
executed that needs access to these registers. Meanwhile, the old
content is left in place with the hope that the current task might not
use these registers at all. This optimization is commonly called lazy
FPU context switching. To make it possible, a processor offers the
ability to toggle the availability of instructions utilizing floating
point and SIMD registers. If the instructions are turned off, any
attempt of executing them will generate a fault.

In this paper, we present an attack that exploits lazy FPU context
switching and allows an adversary to recover the FPU and SIMD register
set of arbitrary processes or VMs. The attack works on processors that
transiently execute FPU or SIMD instructions that follow an instruction
generating the fault indicating the first use of FPU or SIMD
instructions. On operating systems using lazy FPU context switching, the
FPU and SIMD register content of other processes or virtual machines can
then be reconstructed via cache side effects.

With SIMD registers not only being used for cryptographic computation,
but also increasingly for simple operations, such as copying memory, we
argue that lazy FPU context switching is a dangerous optimization that
needs to be turned off in all operating systems, if there is a chance
that they run on affected processors.
\end{abstract}

\hypertarget{introduction}{%
\section{Introduction}\label{introduction}}

As demonstrated by the Meltdown attack {[}4{]}, Intel processors
speculatively execute instructions past the kernel/user privilege check
and the generation of a page fault. Together with the related Spectre
attack {[}3{]}, these attacks were a revelation that microarchitectural
design decisions in processors affect security properties of computing
devices and sparked research into discovering further security-relevant
design issues in current general purpose CPU architectures.

In this paper, we introduce a new information leak vulnerability similar
to Meltdown that affects popular operating systems and hypervisors. We
present practical attacks based on this vulnerability leaking FPU
register state across process and virtual machine boundaries. In
contrast to Meltdown, we exploit the behavior of recent Intel processors
when they encounter a Device Not Available (\texttt{\#NM}) exception
instead of a Page Fault (\texttt{\#PF}) exception. This exception is
used to implement a context switch optimization called \emph{lazy FPU
context switching}. We will demonstrate how speculative instruction
execution can lead to full recovery of the FPU register state of a
victim process using unprivileged local code execution in combination
with this optimization.

The paper is structured as follows. We start with a background section
that gives an overview over FPU state context switching in operating
systems and how the lazy FPU context switching optimization works on
Intel processors. We will briefly revisit architectural details related
to speculative execution on out-of-order processors. Equipped with this
background knowledge, the following section builds the LazyFP attack
from a one-bit leak towards several practical attack variants. We will
evaluate these variants. After assessing the impact of this
vulnerability to Intel AES-NI and discussing mitigation, we review
related work and finally conclude.

\hypertarget{background}{%
\section{Background}\label{background}}

In this section, we give relevant technical background that is necessary
to understand the LazyFP vulnerability.

\hypertarget{the-x87-fpu}{%
\subsection{The x87 FPU}\label{the-x87-fpu}}

The x87 floating-point unit (FPU) is a processor extension with the
original purpose of accelerating mathematical operations on
floating-point numbers. It has its own set of instructions and
registers. It is an integral part of every Intel x86 microprocessor
since the Intel 486DX introduced in 1989. Up to the Intel 486DX, it used
to be an optional external co-processor.

Saving and restoring the FPU state to and from memory, which is required
to implement multitasking, was costly at that time because memory was
slow and limited. Additionally, at that time usually only few
applications actually used the FPU. Switching FPU states on every
context switch, although the FPU is not used by all processes imposes
unnecessary overhead. In order to be able to reduce this overhead, a
control register bit (\texttt{cr0.ts}) was introduced that allowed the
operating system to switch the FPU on and off. While the FPU is turned
off, it keeps its state, but is inaccessible to both user and kernel
code.

The optimization made possible by \texttt{cr0.ts} is called \emph{lazy
FPU context switching}. The core idea is that FPU register state is only
context switched when necessary and left in-place for processes that do
not use the FPU. This way, the operating system avoids the cost of
saving and restoring FPU context where possible.

The operating system tracks which process the current FPU register state
belongs to. This process is called the \emph{FPU owner}. Additionally,
the FPU may be enabled or disabled.

The simplest case is when the FPU is enabled. Here, the current process
is the FPU owner and can freely execute FPU instructions. When the
operating system switches to another process, the FPU is disabled. It
stays disabled until a process attempts to execute an FPU instruction.
As the FPU is disabled, the processor generates an \#NM exception.

What happens in response to the \texttt{\#NM} depends on whether the
current process is the FPU owner. If it is, the FPU register state
belongs to this process and the operating system simply enables the FPU.
The process is again free to execute FPU instructions.

If the current process when receiving the \texttt{\#NM} exception is not
the FPU owner, the operating system enables the FPU, stores the current
FPU register state to the save area for the FPU owner and restores the
FPU register state of the current process. As the FPU register state now
matches the running process, this process is now the FPU owner.

Variations on the above algorithm are possible, but all of them share
the property that the operating system does not have to context switch
the FPU register state when it switches from an FPU-using process to a
process that does not use the FPU and back.

While \emph{lazy FPU context switching} is still widely used in many
operating systems today, its underlying assumption is usually not true
anymore. Starting with the Intel MMX instruction set extension and
continuing with SSE, AVX, and AVX-512, the FPU register state has been
extended with ever larger SIMD registers. The SSE instruction set is
mandatory for 64-bit x86 processors and practically all programs and
libraries use it for various tasks that would be inefficient without it,
such as copying memory.

Even though the larger FPU register state makes it theoretically more
attractive to avoid costly FPU context switches, in the current software
landscape, every process will eventually touch FPU registers and cause a
costly \texttt{\#NM} exception. For this reason, at least the Linux
kernel has switched to eager FPU context switching by default, where FPU
registers are switched as part of the normal context switch and no
\texttt{\#NM} exception is generated in normal operation.

\hypertarget{fpu-handling-in-virtualized-systems}{%
\subsection{FPU Handling in Virtualized
Systems}\label{fpu-handling-in-virtualized-systems}}

A complete description of virtualization is out of scope for this paper.
For the purposes of this paper it is sufficient to understand that with
Intel VT the processor does not take care of switching FPU register
state when transitioning from guest mode to the hypervisor. This leaves
the hypervisor in charge of context switching FPU register state and the
mechanisms are the same as for a non-virtualized operating system. As
such, lazy FPU context switching is also applicable to hypervisors.

\hypertarget{speculative-execution}{%
\subsection{Speculative Execution}\label{speculative-execution}}

Speculative execution is a technique used by modern microprocessors
aimed at increasing the utilization of the processor pipeline. The
processor executes instructions ahead of time, but prevents their
results, such as register updates and memory writes, from becoming
architecturally visible until it was veryfied that they are indeed
supposed to be executed.

During this speculative execution, the processor has incomplete
information. For example, it has to predict the direction of conditional
or indirect jumps. It can also assume that certain instructions do not
generate exceptions. If these ``guesses'' turn out to be correct once
the relevant information is available, the results can be committed to
the architectural state and the respective instruction retires. If an
assumption was invalid, the processor discards all results from its
mis-speculation and resumes execution from the last correctly executed
instruction.

Instructions that execute, but their architectural side-effects are
discarded, are called \emph{transient} instructions. However, these
instructions can change microarchitectural state. Microarchitectural
state is not part of the processor's instruction set architecture. An
example of microarchitectural state is the content of caches.

For an in-depth discussion of speculative execution and how
microarchitectural state can be converted into architectural state, we
refer the reader to the Meltdown paper {[}4{]}.

\hypertarget{intel-tsx}{%
\subsection{Intel TSX}\label{intel-tsx}}

Intel Transactional Synchronization eXtensions (TSX) is the product name
for two x86 instruction set extensions, called Hardware Lock Elision
(HLE) and Restricted Transactional Memory (RTM). HLE is a set of
prefixes that can be added to specific instructions. These prefixes are
backward-compatible so that code that uses them also works on older
hardware.

RTM is an extension adding several instructions to the instruction set
that are used to declare regions of code that should execute as part of
a hardware transactions. Transactions can protect a series of memory
accesses that shall either all succeed together or shall be rolled back
together in case of any error condition or concurrent access by other
threads.

A RTM transaction comprises the region of code that is encapsulated
between a pair of \texttt{xbegin} and \texttt{xend} instructions.
\texttt{xbegin} also provides a mechanism to define an fall-back handler
that is called if the transaction is aborted. \texttt{xabort} can be
used by the executing code to explicitly abort the transaction. In
addition to that, the processor might abort the transaction upon certain
events. These events include, among others, an exception that occurs
during the transaction.

When we refer to Intel TSX in this paper, we refer to RTM specifically.

\hypertarget{leaking-fpu-registers}{%
\section{Leaking FPU Registers}\label{leaking-fpu-registers}}

In this section, we will build several closely related exploits,
starting with a simple one-bit leak of an FPU register and extending it
towards leaking the complete FPU register set.

In the following discussion, we assume that there is a victim and an
attacker process. The victim process contains confidential information
in its FPU register set, such as cryptographic keys. The attacker
process needs no special permissions beyond the ability to execute
arbitrary user code on the same processor core (hardware thread) as the
victim. Mechanisms that achieve co-location are out of scope for this
paper, but in the simplest case random chance helps with this
requirement. Also typical operating systems do not regard the ability to
pin threads to cores as a privileged operation.

\hypertarget{the-one-bit-leak}{%
\subsection{The One-Bit Leak}\label{the-one-bit-leak}}

A simple x86 assembly program for a one-bit leak is shown in Figure
\ref{fpubasic}. It reads the lower-half of the SSE registers
\texttt{xmm0} into the \texttt{rax} general purpose register. It then
masks the lowest bit and shifts it by 6 to be either 0 or 64 depending
on the input value. This calculated offset is used to write to memory.
On a system with 64-byte cache lines, the write operation will touch one
of two cache lines depending on bit 0 of the \texttt{xmm0} register.

\lstset{basicstyle=\ttfamily\footnotesize,breaklines=true,
        morekeywords={movq,xbegin,xabort}, numbers=left}
\lstset{language={[x86masm]Assembler}, xleftmargin=2em,frame=none,framexleftmargin=1.5em}

\begin{figure}[htp]
\begin{lstlisting}
movq rax, xmm0
and  rax, 1
shl  rax, 6
mov  dword [mem + rax], 0
\end{lstlisting}
\caption{The basic building block of the LazyFP attack. The FPU register access
executes speculatively with the previous process' FPU register set. The execution
is retried once the operating system kernel handles the \#NM exception generated
by the first instruction.}
\label{fpubasic}
\end{figure}

Assuming the FPU registers being owned by the victim process due to lazy
FPU context switching, the \texttt{movq} instruction generates a
\texttt{\#NM} fault to indicate to the operating system that the FPU is
disabled. The operating system will transparently handle this fault,
restore the register state of the current process and continue
execution.

The interesting part happens before the processor retires the
\texttt{movq} and generates the \#NM fault. It has already executed the
subsequent instructions speculatively. The architectural changes caused
by these instructions is discarded, but their microarchitectural
footprint in the cache is not.

We thus get a speculative execution of the code with the victim's FPU
register set and the regular execution with the attacker's register set.
Assuming that the attacker set \texttt{xmm0} to zero and flushed
\texttt{mem\ +\ 64} from the processor cache, he can now recover the
victim's bit by probing the access latency of this memory location.

While this attack can be repeated to leak arbitrary bits from arbitrary
registers, in this simple form it is not practical. Each attempt at
leaking needs to be preceded by letting the victim run. This is
necessary to move ownership of the FPU back to the victim, but it also
means the victim will likely change its register content rendering the
results hard to use.

\hypertarget{leaking-the-complete-register-set}{%
\subsection{Leaking the Complete Register
Set}\label{leaking-the-complete-register-set}}

To leak a consistent snapshot of the FPU register set of the victim
without the victim getting a chance to change the state, the attacker
needs a way to suppress the generation of the \texttt{\#NM} exception.

One idea is to suppress the \texttt{\#NM} exception by deliberately
triggering another exception that the attacker can handle before reading
from the FPU registers.

\begin{figure}[htp]
\begin{lstlisting}
  mov  dword [0], 0 ; causes #PF
  movq rax, xmm0
  and  rax, 1
  shl  rax, 6
  mov  dword [mem + rax], 0
\end{lstlisting}
\caption{The FPU register access in the shadow of a page fault. Subsequent
instructions will execute speculatively and their results are discarded when the
faulting instruction retires.}
\label{fpupf}
\end{figure}

In Figure \ref{fpupf}, we show a variant of the exploit that causes a
page fault before the instruction touching the FPU register state is
executed. To recover from the page fault exception instead of being
aborted by the operating system, the attacker needs to configure a
signal handler beforehand.

The attack still proceeds as described earlier with the crucial
difference that the instruction touching FPU state is executed only
speculatively. As such, the exception it causes is never generated.
Instead, the attacker receives a signal that he can handle and continue.
The operating system does not see the \#NM exception and will not
replace the victim's FPU register state from the hardware registers.
This leaves the FPU register state untouched and the attack can continue
by probing further bits.

A downside to this approach is that it puts signal handling in the path
between probing the victim's FPU registers and being able to observe the
cache effects. This introduces noise as the cache line that was pulled
in by speculative execution may be evicted during signal handling.

\hypertarget{suppressing-exceptions-using-intel-tsx}{%
\subsection{Suppressing Exceptions using Intel
TSX}\label{suppressing-exceptions-using-intel-tsx}}

The exception suppression using a page fault works, but it is
heavy-weight. For each leaking attempt, a signal is generated by the
kernel that needs to be handled. As we will see in the evaluation, this
slowness reduces the practicality of the attack. Using a more
light-weight way of suppressing the \texttt{\#NM} exception is therefore
desirable. In recent TSX-capable Intel CPUs, the Restricted
Transactional Memory instruction set extension is useful to achieve
exactly that. RTM transactions abort when exceptions are encountered
during transactional execution.

\begin{figure}[htp]
\begin{lstlisting}
  xbegin abort
  movq rax, xmm0
  and rax, 1
  shl rax, 6
  mov dword [mem + rax], 0
  xabort
abort:
\end{lstlisting}
\caption{The FPU register access is executed inside of a TSX transaction.
The transaction will abort once the \#NM exception is generated, but subsequent instructions
have already executed speculatively.}
\label{fputsx}
\end{figure}

By executing the same simple attack as shown before, but inside a TSX
hardware transaction, the attacker can thus leak one bit without
triggering the operating system's handling of any exception. Example
code is shown in Figure \ref{fputsx}.

\hypertarget{suppressing-exceptions-using-retpoline}{%
\subsection{Suppressing Exceptions using
Retpoline}\label{suppressing-exceptions-using-retpoline}}

Intel TSX is a relatively recent addition (Haswell and onwards) to the
Intel x86 instruction set. This way it limits the applicability of the
attack to recent processors. It is possible to achieve similar exception
suppression using the retpoline {[}5{]} construct.

\begin{figure}[htp]
\begin{lstlisting}
  call set_up_target
capture:
  pause
  jmp capture
set_up_target:
  mov [rsp], destination
  ret
\end{lstlisting}
\caption{The Retpoline construct that is used to prevent the CPU to speculate
past an indirect branch. The CPU will always mispredict the target of the `RET`
instruction and speculatively execute the capture loop.}
\label{origretpoline}
\end{figure}

Retpoline is initially meant as Spectre mitigation. The original
Retpoline construct is given in Figure \ref{origretpoline}. We refer to
the original publication {[}5{]} for details, but the idea is to capture
speculative execution in the pause loop (line 3 and 4) until the CPU
notices the misprediction, discards any speculatively executed
instructions, and continues execution at the desired jump destination.

The misprediction is constructed by exploiting the processor's reliance
on the return stack buffer (RSB) for predicting the target of a
\texttt{ret} instruction. When executing a \texttt{call} instruction,
the processor pushes the return address both onto the architectural
stack in memory and onto the RSB. On a subsequent \texttt{ret}, the
processor can pop a value from the RSB to predict the return target,
while retrieving the actual return address from memory lazily. Retpoline
creates a mismatch between these two values by modifying the return
address on the stack in memory. Speculative execution after \texttt{ret}
follows the address from the RSB. Once the processor has fetched the
actual return value from memory, it will notice the misprediction,
discard any results and continue execution at the address fetched from
memory.

\begin{figure}[htp]
\begin{lstlisting}
  call set_up_target
  movq rax, xmm0
  and rax, 1
  shl rax, 6
  mov dword [mem + rax], 0
capture:
  pause
  jmp capture
set_up_target:
  mov [rsp], destination
  ret
destination:
  ; cache line access
  ; latency probing code...
\end{lstlisting}
\caption{Using the Retpoline construct to speculatively execute the FPU register
access in a reliable way.}
\label{retpolineleak}
\end{figure}

Using a technique described by Wong {[}6{]}, we can repurpose Retpoline
for exception suppression. The example code is shown in Figure
\ref{retpolineleak}.

If we prepend the capture loop in the Retpoline construct with the FPU
register access, the instructions we insert will only be executed
speculatively. The pause loop is not removed entirely because it keeps
the processor from continuing speculative execution into the following
cache line access latency probing code, which would tamper with the
results. Any exception that occurs in the speculative execution path is
never delivered. The result is comparable to the Intel TSX variant, but
applicable to older processors without support for TSX and also, as
shown in the Evaluation, slightly faster.

\hypertarget{evaluation}{%
\section{Evaluation}\label{evaluation}}

The practicality of a LazyFP attack depends on the time it takes to leak
data from the FPU register set. The more time the attacker needs to leak
individual bits, the more likely it is that the operating system
preempts the attacker and the victim gets a chance to continue
executing. At that point, the content of the FPU register set changes.

We thus evaluate the different variants of the attack introduced in the
previous chapter by comparing the time it takes to leak a single AVX
register (256 bits of data). One AVX register is enough to leak an Intel
AES-NI cipher key.

We perform our evaluation on an Intel Core i7-5600U CPU running at 2.60
GHz using FreeBSD 11.1. Victim and attacker processes are pinned to the
same logical CPU.

\begin{table}[t]
\centering
\begin{tabular}{l|rr}
\textbf{Method} & \textbf{Cycles} & \textbf{Eff. Throughput} \\
\hline
Page fault & 359.9K & 0.22 MiB/s \\
Intel TSX & 25.4K & 3.12 MiB/s \\
Retpoline & 24.0K & 3.30 MiB/s \\
\end{tabular}
\caption{The cycle count required to leak a single 256-bit AVX register with different exception suppression methods.}
\label{oneregresults}
\end{table}

In Table \ref{oneregresults}, we see that the page fault method is by
far the slowest of the three variants. To leak a single AVX register, it
takes more than 300000 cycles, or 138 µs. This inefficiency is
problematic, because the typical time slice length in a modern operating
system is as low as 1 ms. This duration is not enough to leak the
complete register file of 16 AVX registers. Larger FPU register sets,
such as those introduced with AVX-512, are even farther out of reach.

While leaking individual registers can still be practical, the
efficiency can be increased by an order of magnitude by using Intel TSX
or Retpoline as exception suppression methods. Both of these variants
are able to leak a complete snapshot of the FPU register state in a
single scheduling time slice, even for large register files, such as
AVX-512.

The attack variants presented in this paper leak one bit per execution
attempt. These can be extended in a straightforward way to leak multiple
bits per execution attempt. Leaking 4 bits per attempt instantly
quadruples the effective throughput of the side-channel.

\hypertarget{impact-on-aes-ni}{%
\section{Impact on AES-NI}\label{impact-on-aes-ni}}

While leaking FPU register state seems not as readily useful to an
adversary compared to reading arbitrary memory as in the Meltdown
attack, the impact on confidential information can be equally
devastating.

A full discussion of the affected processors, operating systems and
cryptographic libraries is out of scope for this paper. For this
information, we refer the reader to the Intel security advisory
\href{https://www.intel.com/content/www/us/en/security-center/advisory/intel-sa-00145.html}{INTEL-SA-00145}.
We would still like to give an impact assessment for Intel AES-NI.

The Intel AES-NI instruction set extension {[}1{]} that is used to
provide hardware acceleration for AES encryption and decryption is a
prime target for a LazyFP-based attack. AES-NI was introduced by Intel
in 2010 with the Westmere microarchitecture and is widely used for
efficient AES implementation.

\begin{figure}
\begin{lstlisting}
  ; the data block is in xmm15.
  ; xmm0-xmm10 hold the round keys
  pxor xmm15, xmm0
  aesenc xmm15, xmm1 ; Round 1
  aesenc xmm15, xmm2 ; Round 2
  aesenc xmm15, xmm3 ; Round 3
  aesenc xmm15, xmm4 ; Round 4
  aesenc xmm15, xmm5 ; Round 5
  aesenc xmm15, xmm6 ; Round 6
  aesenc xmm15, xmm7 ; Round 7
  aesenc xmm15, xmm8 ; Round 8
  aesenc xmm15, xmm9 ; Round 9
  aesenclast xmm15, xmm10 ; Round 10
  ; xmm15 holds the encryption result
\end{lstlisting}
\caption{The example AES-128 decryption sequence as it is given in the AES-NI
documentation. All round keys are kept in SSE registers and are thus within
reach of the LazyFP attack.}
\label{aesnidec}
\end{figure}

Figure \ref{aesnidec} shows the AES-128 decryption code, as it is found
in the official Intel documentation. All round keys that are required to
decrypt the particular data block are kept in SSE registers. SSE
registers are part of the FPU register set and thus the LazyFP
vulnerability puts these into reach of an adversary with the ability to
execute code on the same system, regardless of privileges.

The AES-NI documentation has further examples of AES-NI accelerated AES
encryption and key expansion, which all keep key material in SSE
registers. This is particularly troublesome as the key expansion example
keeps the original cipher key in an SSE register for an extended period
of time.

We conclude that AES-NI, especially if it is used as the Intel
documentation suggests, cannot provide confidentiality on a system that
is affected by the LazyFP vulnerability.

The susceptibility of AES-NI to the LazyFP side-channel is ironic, as it
is designed to defend against earlier timing and cache-based
side-channels. This also means that not using AES-NI on affected systems
is not an effective mitigation, as it re-enables these earlier attacks.

\hypertarget{mitigation}{%
\section{Mitigation}\label{mitigation}}

A long term solution is to abandon the idea of lazy FPU context
switching and switch context eagerly instead. For the majority of
operating systems that do not make such functionality configurable, this
change from lazy to eager context switching can only be implemented by
the operating system vendor.

For operating systems that do have configurable FPU context switching,
an effective mitigation is to manually switch to eager context
switching. When running an affected Linux kernel version newer than
Linux 3.7, this can be achieved by adding ``\texttt{eagerfpu=on}'' to
the Linux kernel boot parameters.

\hypertarget{related-work}{%
\section{Related Work}\label{related-work}}

Our paper touches on the areas of processor microarchitectural
vulnerabilities and operating system design.

Meltdown {[}4{]} and Spectre {[}3{]} are the original classes of
vulnerabilities that introduced the concept of exploiting
microarchitectural state to read sensitive memory locations. Our paper
extends this work by adding the ability to read sensitive information
from the FPU register set.

The Flush+Reload technique {[}7{]} is a key building block to translate
microarchitectural state into architectural state and thus make the
above vulnerabilities and the one described in this paper possible.

Jang et al. {[}2{]} demonstrate a practical timing channel using Intel
TSX. The timing channel is based on timing behavior of TSX aborts due to
page faults. We leverage this insight and utilize the ability of TSX to
mask arbitrary exceptions from the operating system.

Retpoline {[}5{]} is a mitigation for Spectre attacks and is based on
the idea that is possible to reliably create branch mispredictions. We
use a Retpoline-inspired technique as a high-performance exception
suppression and speculative execution steering method.

\hypertarget{conclusion}{%
\section{Conclusion}\label{conclusion}}

We have shown that speculative execution can be used to leak
architecturally inaccessible register state on Intel x86 processors
using the example of the FPU register set. This register set includes
SIMD registers that are widely used for cryptographic purposes. As an
example, we have argued how AES-NI, a popular AES-accelerating
instruction set extension, is significantly weakened when it is executed
on a system vulnerable to LazyFP.

We consider this result as another point in the argument that
microarchitectural design of processors has profound implications for
the security of system software and the isolation properties of a
system.

While short-term mitigations can work around the specific issue
presented in this paper and there are band-aids for similar issues, we
believe that a fundamental shift needs to happen in they way processors
are designed. Security needs to take a front row seat.

\hypertarget{responsible-disclosure}{%
\section{Responsible Disclosure}\label{responsible-disclosure}}

The vulnerability was responsibly disclosed by the authors to Intel in
February 2018 and has been assigned
\href{https://cve.mitre.org/cgi-bin/cvename.cgi?name=CVE-2018-3665}{CVE-2018-3665}.
Intel published the vulnerability as
\href{https://www.intel.com/content/www/us/en/security-center/advisory/intel-sa-00145.html}{INTEL-SA-00145}.
The authors have received a bug bounty from Intel.

\hypertarget{acknowledgements}{%
\section{Acknowledgements}\label{acknowledgements}}

The authors would like to thank Anthony Liguori and Tor Lund-Larsen for
their support in the publication of this paper, and Chad Skinner for
managing our relations with Intel. We would also like to thank Werner
Haas and Jacek Galowicz for reviewing this paper and providing valuable
feedback. Any remaining errors have been added to the paper by the
authors.

\hypertarget{references}{%
\section*{References}\label{references}}
\addcontentsline{toc}{section}{References}

\hypertarget{refs}{}
\leavevmode\hypertarget{ref-aesni}{}%
{[}1{]} Gueron, S. 2012. Intel Advanced Encryption Standard (AES) New
Instructions Set.
\url{https://software.intel.com/sites/default/files/article/165683/aes-wp-2012-09-22-v01.pdf}.

\leavevmode\hypertarget{ref-Jang2016}{}%
{[}2{]} Jang, Y., Lee, S. and Kim, T. 2016. Breaking kernel address
space layout randomization with Intel TSX. \emph{Proceedings of the 2016
acm sigsac conference on computer and communications security} (New
York, NY, USA, 2016), 380--392.

\leavevmode\hypertarget{ref-Kocher2018spectre}{}%
{[}3{]} Kocher, P., Genkin, D., Gruss, D., Haas, W., Hamburg, M., Lipp,
M., Mangard, S., Prescher, T., Schwarz, M. and Yarom, Y. 2018. Spectre
attacks: Exploiting speculative execution. \emph{ArXiv e-prints}. (Jan.
2018).

\leavevmode\hypertarget{ref-Lipp2018meltdown}{}%
{[}4{]} Lipp, M., Schwarz, M., Gruss, D., Prescher, T., Haas, W.,
Mangard, S., Kocher, P., Genkin, D., Yarom, Y. and Hamburg, M. 2018.
Meltdown. \emph{ArXiv e-prints}. (Jan. 2018).

\leavevmode\hypertarget{ref-Retpoline}{}%
{[}5{]} Retpoline: A software construct for preventing
branch-target-injection: 2018.
\emph{\url{https://support.google.com/faqs/answer/7625886}}. Accessed:
2018-06-09.

\leavevmode\hypertarget{ref-Wong2018}{}%
{[}6{]} The microarchitecture behind Meltdown: 2018.
\emph{\url{http://blog.stuffedcow.net/2018/05/meltdown-microarchitecture/}}.
Accessed: 2018-06-10.

\leavevmode\hypertarget{ref-Yarom2014FlushReload}{}%
{[}7{]} Yarom, Y. and Falkner, K. 2014. FLUSH+RELOAD: A high resolution,
low noise, L3 cache side-channel attack. \emph{23rd USENIX security
symposium (USENIX security 14)} (San Diego, CA, 2014), 719--732.

\end{document}